\documentclass[prd,showpacs]{revtex4}
\usepackage{epsfig}
\usepackage{amsmath}
\usepackage{amssymb}

\begin{document}
\title{Branching Ratio and Polarization of $B\to\rho (\omega)\rho(\omega)$
Decays\\ in Perturbative QCD Approach }
\author{
 Ying Li\footnote{liying@mail.ihep.ac.cn}, Cai-Dian L\"u}
\affiliation{\it \small  CCAST (World Laboratory), P.O. Box 8730,
Beijing 100080, China;} \affiliation
 {\it \small Institute of High
Energy Physics, P.O.Box 918(4), Beijing 100049,
China\footnote{Mail address}}
\begin{abstract}
In this work, we calculate the branching ratios, polarization
fractions and CP asymmetry parameters of decay modes $B\to\rho
(\omega)\rho (\omega)$ in the  perturbative QCD approach, which is
based on $\mathbf{k}_T$ factorization. After calculation, we find
the branching ratios of $B^0 \to \rho^+ \rho^-$, $B^+ \to \rho^+
\rho^0$ and $B^+ \to \rho^+ \omega$ are at the order of $10^{-5}$,
and their longitudinal polarization fractions are more than
$90\%$. The above results agree with BaBar's measurements. We also
calculate the branching ratios and polarization fractions of $B^0
\to \rho^0 \rho^0$, $B^0 \to \rho^0 \omega$ and $B^0 \to \omega
\omega$ decays. We find that their longitudinal polarization
fractions are suppressed to 60-80\% due to a small color
suppressed tree contribution. The dominant penguin and
non-factorization tree contributions equally contribute to the
longitudinal and transverse polarization, which will be tested in
the future experiments. We predict the CP asymmetry of $B^0 \to
\rho^+ \rho^-$ and $B^+ \to \rho^+ \rho^0$, which will be measured
in $B$ factories.
\end{abstract}
 \maketitle

\section{Introduction}

The study of exclusive non-leptonic weak decays of $B$ mesons
provides not only good opportunities for testing the Standard
Model (SM) but also powerful means for probing different new
physics scenarios beyond the SM. The mechanism of two body $B$
decay is still not quite clear, although many physicists devote to
this field. The hadronic effects must be important while a
reliable calculation of these effects is very difficult
\cite{hep-ph/0411373}. Starting from factorization hypothesis
\cite{BSW}, many approaches have been built to explain the
existing data and made some progress such as generalized
factorization \cite{hep-ph/9804363}, QCD factorization (BBNS)
\cite{hep-ph/0308039 }, perturbative QCD Approach (PQCD)
\cite{hep-ph/9411308}, and soft-collinear effective theory (SCET)
\cite{hep-ph/0109045}. These approaches separately explained many
of the $B \to PP$ and $B \to PV$ decays though some flaws existed
in different approaches.

Recently, $B \to VV$ decays such as $B\to \phi K^{\ast}$
\cite{hep-ph/0204166}, $B\to \rho K^{\ast}$ \cite{lu-shen}, have
aroused  many interests of physicists. It is known that both
longitudinal and transverse polarization states are possible in $B
\to V V$ decay modes. So, the theoretical analysis of $B \to V V$
is more complicated than $B \to P P$ and $B \to P V$. The
predictions of those decays' polarization fractions according to
the naive factorization do not agree with the experimental
results, although many ideas
\cite{hep-ph/0405134,hep-ph/0406162,hep-ph/0409286} have been
proposed to explain this phenomenon. Some people think that it is
a signal of new physics \cite{hep-ph/0411211,hep-ph/0310229}. Very
recently, both BaBar and Belle have measured the branching ratios
and polarizations of the decays $B^0\to\rho^+\rho^-$ and
$B^+\to\rho^+\rho^0$
 \cite{hep-ex/0307026,hep-ex/0311017,hep-ex/0306007,hep-ex/0503049},
some decay modes have very large branching ratios. The
longitudinal polarization fractions are also very large, which are
different from that of $B \to \phi K^*$.

In this paper, we will study the branching ratios, polarization
fractions and CP violation parameters of $B\to\rho
(\omega)\rho(\omega)$ decays in the PQCD approach. At the rest
frame of $B$ meson, the $B$ meson decays to light vector mesons
with large momentum.  Because the two light mesons move fast back
to back, they have small chance to exchange soft particles,
therefore the soft final state interaction may not be important. A
hard gluon emitted from the four quark operator kicks the light
slow spectator quark in $B$ meson with large momentum transfer to
form a fast moving final state meson. Therefore, the short
distance hard process dominates this decay amplitude. In this
factorization theorem, decay amplitude is written as the
convolution of the corresponding hard parts with universal meson
distribution amplitudes, which describe non-perturbative hadronic
process of the decay. Because the Sudakov effect from $k_T$ and
threshold resummation \cite{hep-ph/0102013}, the end point
singularities do not appear.

This paper is organized as follows. In Section~\ref{section:2}, we
give some ingredients of the basic formalism. The numerical
results for branching ratios and CP asymmetry are given in
Section~\ref{sc:result} and \ref{sc:CP} respectively. We summarize
our work at Section \ref{sc:sum}.

\section{Formalism}\label{section:2}
The recently developed PQCD approach is based on $k_T$
factorization scheme, where three energy scales are involved
\cite{hep-ph/9411308}. The hard dynamics is characterized by
$\sqrt{m_B \Lambda_{QCD}}$, which is to be perturbatively
calculated in PQCD. The harder dynamics is from $m_W$ scale to
$m_B$ scale described by renormalization group equation for the
four quark operators. The dynamics below $\sqrt{m_B
\Lambda_{QCD}}$ is soft, which is described by the meson wave
functions. The soft dynamics is not perturbative but universal for
all channels. Based on this factorization, the $B \to \rho \rho$
decay amplitude is written as the following factorizing formula
 \cite{hep-ph/9607214},
\begin{eqnarray}
\mathcal{M} &\sim& \int\!\! dx_1 dx_2 dx_3 b_1db_1 b_2db_2
b_3db_3\  \nonumber \\
 &&\times \mathrm{Tr} \bigl[ C (t) \Phi_B (x_1,b_1)
\Phi_{\rho} (x_2,b_2) \Phi_{\rho} (x_3,b_3) H (x_i,b_i,t)S_t
(x_i)e^{-S (t)} \bigr], \label{eq:convolution1}
\end{eqnarray}
where $\mathrm{Tr}$ denotes the trace over Dirac and color
indices. $C (t)$ is Wilson coefficient of the four quark operator
which results from the radiative corrections at short distance.
The wave function $\Phi_{M}$  absorbs non-perturbative dynamics of
the process, which is process independent. The hard part $H$ is
rather process-dependent and can be calculated in perturbative
approach. $t$ is  chosen as the largest energy scale in the hard
part, to kill the largest logarithm. The jet function $S_t (x_i)$,
called threshold resummation, comes from the resummation  of the
double logarithms $\ln^2x_i$. The Sudakov form factor $S (t)$, is
 from the resummation of double logarithms $\ln^2 Qb$
\cite{hep-ph/9411308,hep-ph/9607214}.

\subsection{Wave Function}
In this paper, we use the light-cone coordinates to describe the
four dimension momentum as $ (p^+,p^-,p^{\perp})$. The $B$ meson
is treated as a heavy-light system, whose wave function is defined
as:
\begin{eqnarray}
\Phi^{ ({\rm in})}_{B,\alpha\beta,ij} &\equiv& \langle 0 |{\bar
b}_{\beta j}(0)d_{\alpha i}(z) | B(p) \rangle
\nonumber\\
&=& \frac{i \delta_{ij}}{\sqrt{2N_c}}\int dx
d^2{\boldsymbol{k}_{T}}
 e^{-i(xp^-z^+-{\boldsymbol{k}_{T}}{\boldsymbol{z}_T})}
\left[ (\not p +M_B)\gamma_5
 \phi_B(x,\boldsymbol{k}_{T})
\right]_{\alpha\beta} \;,
\end{eqnarray}
where the indices $\alpha$ and $\beta$ are spin indices, $i$ and
$j$ are color indices, and $N_c=3$ is the color factor. The
distribution amplitude $\phi_B$ is normalized as
\begin{eqnarray}
\int_0^1 dx_{1}\phi_B (x_{1},b_1=0)&=&\frac{f_B}{2\sqrt{2N_c}}\;,
\label{eq:bnor}
\end{eqnarray}
where $b_1$ is the conjugate space coordinate of  transverse
momentum $k_T$, and $f_B$ is the decay constant of the $B$ meson.
In this study, we use the model function
\begin{eqnarray}
\phi_B (x,b) &=& N_B x^2
 (1-x)^2\exp\left[-\frac{1}{2}\left (\frac{xM_B}{\omega_{B}}\right)^2
-\frac{\omega_{B}^2 b^2}{2}\right] \;,
\end{eqnarray}
where $N_B$ is the normalization constant. We use $\omega_B=0.4$
GeV, which is determined by the calculation of form factors and
other well known decay modes \cite{hep-ph/9411308}.

As a light-light system, The $\rho^-$ meson wave function of the
longitudinal part is given  by  \cite{hep-ph/0412079}
\begin{eqnarray}
\Phi_{\rho^-,\alpha\beta,ij} &\equiv& \langle \rho (p,\epsilon_L)
| \bar d_{\beta j} (z) u_{\alpha i} (0) | 0 \rangle
\nonumber\\
& & \hspace{-20mm} = \frac{\delta_{ij}}{\sqrt{2N_c}}\int^1_0dx
e^{ixp\cdot z} \left[ m_\rho\not\epsilon_L
 \phi_\rho (x)+
\not\epsilon_L \not p
 \phi_\rho^t (x)
+ m_\rho
 \phi_\rho^s (x)
\right]_{\alpha\beta}.
\end{eqnarray}
The first term in the above equation is the leading twist wave
function (twist-2), while the others are sub-leading twist
(twist-3) wave functions. The $\rho$ meson can also be
transversely polarized, and its wave function is then
\begin{eqnarray}
<\rho^- (p,\epsilon_T)|\overline{d}_{\beta j} (z)u_{\alpha i}
(0)|0> &=& \frac{\delta_{ij}}{\sqrt{2N_c}}\int_0^1 dx e^{ixp\cdot
z} \left\{ \not \epsilon_T \left[\not p
\phi_{\rho}^T  (x) + m_{\rho} \phi_{\rho}^v  (x) \right] \nonumber\right.\\
&&\left.+ \frac{m_{\rho}}{p\cdot n}
 i\epsilon_{T\mu\nu\rho\sigma}\gamma_5\gamma^\mu \epsilon^\nu p^\rho
n^\sigma\phi_{\rho}^a  (x)\right\},
\end{eqnarray}
where $n$ is the moving direction of $\rho$ particle. Here the
leading twist wave function for the transversely polarized
 $\rho$ meson is the first term which is proportional to $\phi_{\rho}^T$.

 The distribution amplitudes of $\rho$ meson, $\phi_{\rho}$, $\phi^t_{\rho}$,
 $\phi^s_{\rho}$, $\phi^T_{\rho}$, $\phi^v_{\rho}$, and $\phi^a_{\rho}$,
 are calculated using light-cone QCD sum rule \cite{hep-ph/0412079}:
 \begin{eqnarray}
\phi_\rho (x)&=&\frac{3f_\rho}{\sqrt{2N_c}} x (1-x)\left[1+
0.18C_2^{3/2} (2x-1)\right]\;,
\label{pwr}\\
\phi_{\rho}^t (x)&=&\frac{f^T_{\rho}}{2\sqrt{2N_c}} \left\{3
(2x-1)^2+0.3 (2x-1)^2[5 (2x-1)^2-3]\right.
\nonumber \\
& &\left.+0.21[3-30 (2x-1)^2+35 (2x-1)^4]\right\}\;,
\label{pwt}\\
\phi_{\rho}^s (x) &=&\frac{3f_\rho^T}{2\sqrt{2N_{c}}}
 (1-2x)\left[1+0.76 (10x^2-10x+1)\right]\;,
\label{pws}\\
\phi_\rho^T (x)&=&\frac{3f_\rho^T}{\sqrt{2N_c}} x (1-x)\left[1+
0.2C_2^{3/2} (2x-1)\right]\;,
\label{pwft}\\
\phi_{\rho}^v (x)&=&\frac{f_{\rho}}{2\sqrt{2N_c}}
\bigg\{\frac{3}{4}[1+ (2x-1)^2]+0.24[3 (2x-1)^2-1]
\nonumber \\
& &+0.12[3-30 (2x-1)^2+35 (2x-1)^4]\bigg\}\;,
\label{pwv}\\
\phi_{\rho}^a (x) &=&\frac{3f_\rho}{4\sqrt{2N_{c}}}
 (1-2x)\left[1+0.93 (10x^2-10x+1)\right]\;, \label{pwa}
\end{eqnarray}
with the Gegenbauer polynomials,
\begin{eqnarray}
& &C_2^{1/2} (t)=\frac{1}{2} (3t^2-1)\;,\;\;\; C_4^{1/2}
(t)=\frac{1}{8} (35 t^4 -30 t^2 +3)\;,
\nonumber\\
& &C_2^{3/2} (t)=\frac{3}{2} (5t^2-1)\;,\;\;\; C_4^{3/2}
(t)=\frac{15}{8} (21 t^4 -14 t^2 +1) \;.
\end{eqnarray}

\subsection{Perturbative calculations}\label{sc:fm}

For decay $B\to \rho\rho$, the related effective Hamiltonian is
given by  \cite{hep-ph/9512380}
\begin{equation}
 H_\mathrm{eff} = \frac{G_F}{\sqrt{2}}\left\{ V_{ud}V_{ub}^* \left[
C_1 (\mu) O_1 (\mu) + C_2 (\mu) O_2 (\mu)
\right]-V_{tb}^*V_{td}\sum_{i=3}^{10}C_i (\mu) O_i
(\mu)\right\},\label{hami}
\end{equation}
 where $C_{i} (\mu) (i=1,\cdots,10)$ are Wilson coefficients at the
  renormalization scale $\mu$ and the four quark operators $O_{i} (i=1,\cdots,10)$ are
\begin{equation}\begin{array}{ll}
  O_1 =  (\bar{b}_iu_j)_{V-A} (\bar{u}_jd_i)_{V-A},  &
  O_2 =  (\bar{b}_iu_i)_{V-A}  (\bar{u}_jd_j)_{V-A},  \\
  O_3 =  (\bar{b}_id_i)_{V-A}\sum_{q}  (\bar{q}_jq_j)_{V-A},  &
  O_4 =  (\bar{b}_id_j)_{V-A}\sum_{q}  (\bar{q}_jq_i)_{V-A}, \\
  O_5 =  (\bar{b}_id_i)_{V-A}\sum_{q}  (\bar{q}_jq_j)_{V+A},  &
  O_6 =  (\bar{b}_id_j)_{V-A} \sum_{q}  (\bar{q}_jq_i)_{V+A}, \\
  O_7 = \frac{3}{2} (\bar{b}_id_i)_{V-A} \sum_{q}
   e_q (\bar{q}_jq_j)_{V+A},   &
   O_8 = \frac{3}{2} (\bar{b}_id_j)_{V-A}\sum_{q} e_q
   (\bar{q}_jq_i)_{V+A}, \\
  O_9 = \frac{3}{2} (\bar{b}_id_i)_{V-A}\sum_{q}
  e_q (\bar{q}_jq_j)_{V-A}, &
   O_{10} = \frac{3}{2} (\bar{b}_id_j)_{V-A}\sum_{q}
 e_q (\bar{q}_jq_i)_{V-A}. \label{eq:effectiv}
 \end{array}
\end{equation}
Here $i$ and $j$ are $SU(3)$ color indices; the sum over $q$ runs
over the quark fields that are active at the scale $\mu=O
(m_{b})$, i.e., $q\in \{u,d,s,c,b\}$. Operators $O_{1}, O_{2}$
come from tree level interaction, while $O_{3}, O_{4}, O_{5},
O_{6}$ are QCD-penguin operators and $O_{7}, O_{8}, O_{9}, O_{10}$
come from electroweak-penguins.

\begin{figure}[htb]
\begin{center}
\includegraphics[scale=0.6]{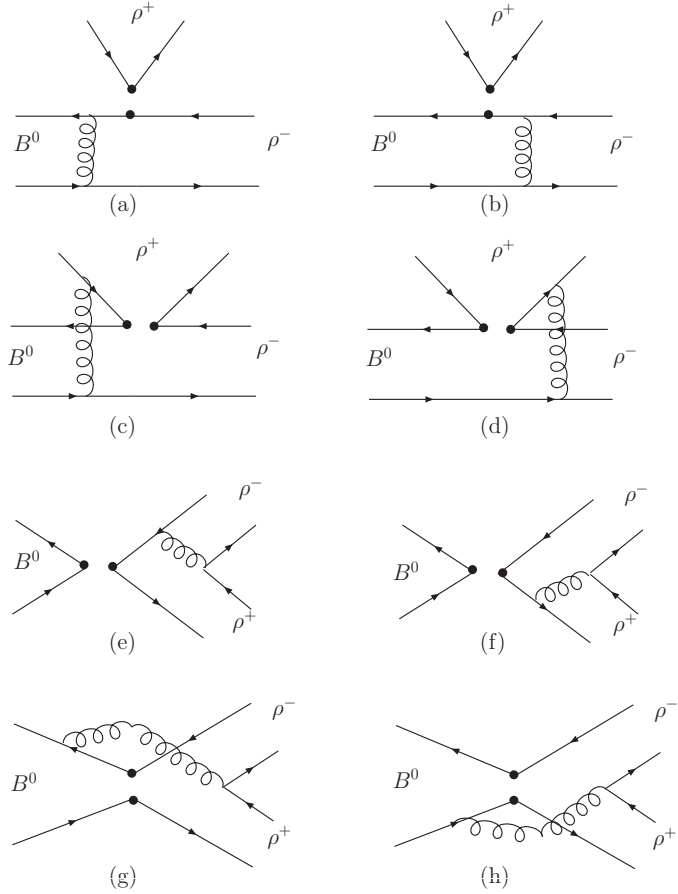}
\caption{The leading order Feynman diagrams for $B \to \rho
\rho$.} \label{fig1}
\end{center}
\end{figure}

Similar to the $B \to \pi \pi$ decays \cite{hep-ph/9411308}, there
are eight types of Feynman diagrams contributing to $B \to \rho^+
\rho^-$ decay mode at leading order, which are shown in
Fig.~\ref{fig1}.
 They involve two types: the emission and annihilation
topologies. Each type is classified into factorizable diagrams,
where hard gluon connects the quarks in the same meson, and
non-factorizable diagrams, where hard gluon attaches the quarks in
two different mesons. Through calculating these diagrams, we can
get the amplitudes $M_H$, where $H=L, N, T$ standing for the
longitudinal and two transverse polarizations. Because these
diagrams are the same as those of $B \to \phi K^*$
\cite{hep-ph/0204166} and $B \to K^*K^*$ \cite{hep-ph/0504187},
the formulas of $B \to \rho \rho$ are similar to those of $B \to
\phi K^*$ or $B \to K^*K^*$. We just need to replace some
corresponding wave functions, Wilson coefficients and
corresponding parameters. So we don't present the detailed
formulas in this paper.

\section{Numerical Results for Branching Ratios and Polarizations}\label{sc:result}

In our calculation, some parameters such as meson mass, decay
constants, the CKM matrix elements and the lifetime of $B$ meson
 \cite{pdg} are given in Table~\ref{para}.

\begin{table}[htb]
\caption{Parameters we used in numerical calculation \cite{pdg}}
\label{para}
\begin{center}
\begin{tabular}[t]{r|cc}
 \hline     \hline
 Mass      &$m_{B^0}=5.28\mbox{GeV}$ & $m_{B^+}=5.28 \mbox{GeV} $ \\
     &$m_{\rho}=0.77 \mbox{GeV}$  &$m_{\omega}=0.78\mbox{GeV}$  \\
 \hline
 \hline
 Decay    & $f_B =196 \mbox{ MeV}$ &$f_{\rho}=f_{\omega}=200 \mbox{MeV}$  \\
 Constants& $f_{\rho}^{\perp}=f_{\omega}^{\perp}=160 \mbox{MeV}$   \\
 \hline
 \hline
 CKM   & $|V_{ud}|= 0.9745$ & $|V_{ub}|= 0.042$ \ \\
       & $|V_{td}|= 0.0025$ & $|V_{tb}|= 0.999$ \ \\
 \hline
 \hline
Lifetime  & $\tau_{B^0}=1.54\times 10^{-12}\mbox{ s}$
\         & $\tau_{B^+}=1.67\times 10^{-12}\mbox{s}$ \ \\
 \hline
\end{tabular}
\end{center}
\end{table}

Taking $B^0 \to \rho^+ \rho^- $ as an example, we know that
($H=L,N,T$):
\begin{eqnarray}\label{M_H}
M_H &=& V_{ub}^*V_{ud}T_H-V_{tb}^*V_{td}P_H \nonumber\\
    &=& V_{ub}^*V_{ud}T_H  (1+z_He^{i (-\alpha+\delta_H)}).
\end{eqnarray}
with definition: CKM phase angle
$\alpha=\arg\left[-\frac{V_{tb}V_{td}^*}{V_{ub}V_{ud}^*}\right]$
and $z_H=\left|\frac{V_{tb}V_{td}^*}{V_{ub}V_{ud}^*}\right|
\left|\frac{P_H}{T_H}\right|$.  The strong phase $\delta_H$ and
ratio $z_H$ of tree $ (T)$ and penguin $ (P)$ are calculated in
PQCD approach. In PQCD approach, the strong phases come from the
non-factorizable diagrams and annihilation type diagrams because
quarks and gluons can be on mass shell. Numerical analysis also
shows that the main contribution to the relative strong phase
$\delta_H$ comes from the penguin annihilation diagrams. $B$ meson
annihilates into $q \bar q$ quark pair and then decays to
$\rho\rho$ final states \cite{hep-ph/0411373,hep-ph/0409317}. In
hadronic picture, the intermediate $q \bar q$ quark pair
represents a number of resonance states, which implies final state
interaction. These diagrams also make the contribution of penguin
diagrams more important than previously expected.

In Table~\ref{ratio}, we show the numerical results of each
diagram in $B^0 \to \rho^+ \rho^-$ decay mode. From this table, we
find that the most important contribution (about $95\%$) comes
 from the first two factorizable emission diagrams Fig.\ref{fig1}
(a) and (b), especially for the longitudinal part. But the first
two diagrams can not contribute to the relative strong phases. The
main source of strong phases are from the annihilation diagrams,
especially penguin diagrams of  Fig.\ref{fig1} (e)(f) and
 (g)(h). We can calculate that the strong phase for each
 polarization is $\delta_L=13.6^{\circ}$, $\delta_N=42^{\circ}$, and
$\delta_T=39^{\circ}$.

\begin{table}[htb]
\caption{Polarization amplitudes $ ( 10^{ -3}\mathrm{GeV})$ of
different diagrams in $B^0 \to \rho^+ \rho^-$ decay} \label{ratio}
\begin{center}
\begin{tabular}[t]{r|c|c|c|c}
 \hline     \hline
 Decay mode      & (a) and (b)&  (c) and (d) & (e) and (f)&  (g) and (h)  \\
 \hline

  $ L (T) $    & $77$ & $-2.4+ 0.6 i $ &$     0     $ & $-1.4-3.4 i$ \\
  \hline
  $ L (P) $    & $-3.1$ & $ 0.14+ 0.03 i $ &$3.0-1.7 i$ & $0.39 + 0.57 i$ \\
  \hline
  $ N (T) $    & $ 8.7$ & $1.3-0.05 i   $ &$    0      $ & $0.04-0.09 i$\\
 \hline
  $ N (P) $    & $-0.34$ & $-0.03+0.007 i $ &$1.6+0.8i$ & $-0.002+0.009 i$\\
 \hline
  $ T (T) $    & $17$ & $2.7-0.004 i  $ &$0.04+0.01 i$ & $0.002-0.008i$\\
 \hline
  $ T (P) $    & $-1.8$ & $-0.07+0.02 i $  &$3.2+1.7 i$ & $0.004+0.004i $\\
 \hline
 \hline
\end{tabular}
\end{center}
\end{table}

\begin{figure}[htb]
\begin{center}
\includegraphics[scale=0.6]{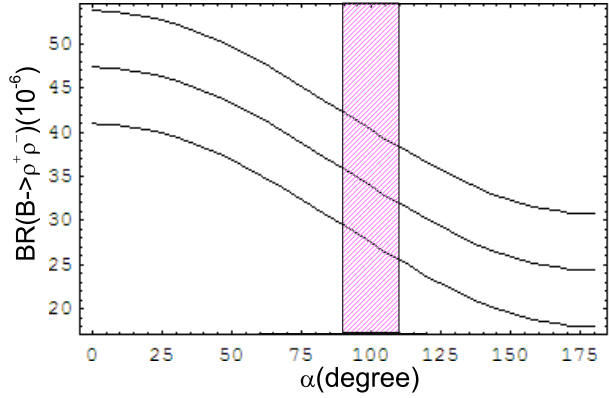}
\caption{Average branching ratio with theoretical uncertainty of
$B^0 \to \rho^+ \rho^-$ as  a function of CKM angle $\alpha$,
where the shaded band shows the 1 $\sigma$  constraint for
$\alpha$} \label{BR+-}
\end{center}
\end{figure}

In the same way, we can get the formula for the charged conjugate
decay $\bar B^0 \to \rho^+ \rho^- $:
\begin{eqnarray}\label{M_H_2}
\bar M_H = V_{ub}V_{ud}^*T_H  (1+z_He^{i (\alpha+\delta_H)}).
\end{eqnarray}
Therefore, the averaged branching ratio for $ B \to \rho^+ \rho^-
$ is:
\begin{eqnarray}\label {M_H_3}
\mathcal{M}^2_H\propto
   |V_{ub}^*V_{ud}T_H| ^2  (1+2z_H\cos\alpha\cos\delta_H+z_H^2).
   \label{18}
\end{eqnarray}
Here, we notice the branching ratio is a function of $\cos
\alpha$. This $\cos \alpha$ behavior of the branching ratio is
shown in Fig.\ref{BR+-}. In principle, we can determine angle
$\alpha$ through eq.(\ref{18}). However, the uncertainty of theory
is so large (also shown in Fig.\ref{BR+-}) to make it unrealistic.
Firstly, the major uncertainty comes from higher order correction.
In the calculation of $B \to K \pi$ \cite{lipre}, the results show
that the next-to-leading order contribution can give about
$15\%-20\%$ correction to leading order. Secondly, the wave
functions which describe the hadronic process of the meson are not
known precisely, especially for the heavy $B$ meson. Using the
existing data of other channels such as $B\to \pi l\nu$
\cite{hep-ph/9409313}, $B\to D \pi$ \cite{hep-ph/0305335}, $B \to
K \pi, \pi\pi$ \cite{hep-ph/9411308}, {\it etc}, we can fit the
$B$ meson wave function parameter $\omega_B=0.4\pm 0.1$. Another
uncertainty comes from parameter $c$ of threshold resummation
\footnote{The formula of threshold resummation
 \cite{hep-ph/0102013}
$S_t (x)=\frac{2^{1+2c}\Gamma (3/2+c)}{\sqrt{\pi}\Gamma (1+c)} [x
(1-x)]^c$}, and it varies from $0.3$ to $0.4$. In leading order,
considering the uncertainty taken by $\omega$ and $c$, we give the
branching ratios and polarization fractions in Table \ref{results}
together with averaging experimental measurements
\cite{hep-ex/0307026,hep-ex/0311017,hep-ex/0306007,hep-ex/0503049}.

There are still many other  parameters existed such as decay
constants, CKM elements, and we will not discuss the uncertainty
here. The polarization fractions of these decay modes are not
sensitive to the above parameters, because they mainly give an
overall change of all polarization amplitudes, not to the
individual noes. From our calculation, we find that these
polarization fractions are sensitive to the distribution
amplitudes of vector mesons. However, the distribution amplitudes
we used are results from Light-Cone Sum
Rules\cite{hep-ph/0412079}, which are difficult to change. Anyway,
20\% uncertainty from the meson distribution amplitudes for the
polarization fractions are possible. The range of CKM angle
$\alpha$ has been well constrained as $
\alpha=(98^{+6.1}_{-5.6})^\circ$ \cite{hep-ph/0511125}, so that
its small uncertainty affects very little on the branching ratios.

\begin{table}[htb]
\caption{Branching ratios and polarizations fractions of $B\to\rho
(\omega)\rho (\omega)$ decays from theory and experiments
\cite{hep-ex/0307026,hep-ex/0311017,hep-ex/0306007,hep-ex/0503049}.
In our results, the uncertainties come from $\omega_B$ and $c$
respectively.} \label{results}
\begin{center}
\begin{tabular}[t]{c|c|c|c|c|c|c|}
\hline \hline
 & \multicolumn{2}{c|}{ BR $ (10^{-6})$ }
 & \multicolumn{2}{c|}{ $f_L (\%)$}&   &
 \\
Channel & Theory &  Exp. &  Theory& Exp. &
$f_{\parallel }(\%)$ & $f_{\perp} (\%)$  \\
\hline $B^0 \to \rho^+ \rho^-   $  & $35 \pm5\pm4$          &
$30\pm6$
 & 94 &$96^{+4}_{-7}$& 3     & 3            \\
$B^+ \to \rho^+ \rho^0   $  & $17 \pm2\pm1$          &   $26.4^{+6.1}_{-6.4}$
 & 94 &$99\pm5$  & 4     & 2            \\
$B^+ \to \rho^+ \omega   $  & $19 \pm2\pm1$           & $12.6^{+4.1}_{-3.8}$
 & 97 &$88^{+12}_{-15}$  & 1.5   & 1.5          \\
$B^0 \to \rho^0 \rho^0   $  & $0.9  \pm0.1\pm0.1$         & $<1.1$
   & 60 & - & 22    & 18           \\
$B^0 \to \rho^0 \omega   $  & $1.9 \pm0.2\pm0.2$         & $<3.3$
               & 87 & - & 6.5   & 6.5          \\
$B^0 \to \omega \omega   $  & $1.2\pm0.2\pm0.2 $         &  $<19$
            & 82 & - & 9     & 9            \\
\hline \hline
\end{tabular}
\end{center}
\end{table}

 From above results and Table~\ref{results}, some discussions are
in order:
\begin{itemize}
     \item For simplicity, we set that the $\rho^0$, $\omega$
          have same mass, decay constant and
          distribution amplitude. In quark model, the $\rho^0$
          meson is $\frac{1}{\sqrt{2}} (u\bar u-d\bar d)$, while
          $\omega$ is $\frac{1}{\sqrt{2}} (u\bar u+d\bar d)$. The
          difference comes from the sign of $d \bar d$, which only
          appears in penguin operators, so their difference should
          be relatively small.
    \item For the tree dominant decays, most of the contribution to branching ratio
          comes from factorizable spectator diagram  (a) and  (b), which
          are the diagrams contribute to
           the $B \to \rho$ form factor.
           For example in decay mode $B^0 \to \rho^+ \rho^- $,
            the dominant Wilson coefficients are $C_2+C_1/3$  (order of 1) in
          tree level, which is supported by numerical results. The decay $B^+ \to \rho^+
          \omega$ and $B^+ \to \rho^+
          \rho^0$ have the similar situation. Their branching
          ratios are all at the order $10^{-5}$.
    \item For decay $B \to \rho^0 \rho^0$, the Wilson coefficient
          is $C_1+C_2/3$ in tree level, which is color suppressed.
           In this work, we only calculate the
          leading order diagrams, and did not calculate the higher
          order corrections. So, the Wilson coefficients we used
          are leading order results in order to keep consistency. In
          leading order, the sign of $C_2$ is positive while the
          sign of $C_1$ is negative, which can cancel  each other mostly.
          Thus
          the branching ratio of $B\to \rho ^0 \rho ^0$ is rather
          small. If considering next to leading order corrections,
          the sign of $C_1+C_2/3$ may change to positive, so the
          branching ratio may become larger. This decay should be
          more sensitive to next leading order contribution. This
          is similar to the argument of $B^0 \to \pi^0 \pi^0$
          decay and $B^0 \to \rho^0 \omega$,
          $\omega\omega$.
    \item Comparing  our results with experiments (world average), we find both branching ratios and
    polarizations
          agree well with only one exception: $B^+ \to \rho^+ \rho^0$.
In fact, this is due to a large branching fraction measured by
Belle \cite{hep-ex/0306007},
\begin{eqnarray}
          \mathbf{BR} (B^+ \to \rho^+ \rho^0  )= (31.7\pm7.1^{+3.8}_{-6.7})\times
           10^{-6},
          \end{eqnarray}
          which does not overlap with
          BaBar's data \cite{hep-ex/0307026,hep-ex/0311017,hep-ex/0503049}
          \begin{eqnarray}
           &\mathbf{BR} (B^+ \to \rho^+ \rho^0  )= (22.5^{+5.7}_{-5.4}\pm5.8)\times
           10^{-6}.
          \end{eqnarray}
 We are waiting for the consistent  results from two experimental groups.
          As for the color suppressed $B^0 \to \rho^0 \rho^0 $, $B^0 \to \rho^0
          \omega$ and $B^0 \to \omega\omega$, there are only upper
          limits now, and our results are still below the upper limits.
    \item In Ref.~\cite{hep-ph/0411211,hep-ph/0104090,hep-ph/0507122},
          these decay modes have been calculated
          in QCD factorization approach. For $B^0 \to \rho^+ \rho^-$, the
          branching ratio they predicted is a bit larger than the
          experimental data, because the form factor they used is $V^{B\to
          \rho}=0.338$. In PQCD approach \cite{hep-ph/0212373}, this form factor is about $0.318$,
           so our results is smaller than theirs. Similar to above
          decay, our results in decay $B^+ \to \rho^+ \rho^0$ is also
          smaller than results in QCD factorization approach for the same
          reason. For decay modes $B^0 \to \rho^0 \rho^0 $, $B^0 \to \rho^0
          \omega$ and $B^0 \to \omega\omega$, our results are much
          larger than theirs because the annihilation diagrams
          play very important role, and these parts
          cannot be calculated
          directly in QCD factorization approach.
    \item Both dominant by color enhanced tree contribution, we can see that the branching
          ratio of $B^0 \to \rho^+ \rho^-$ is about two times of that of
          $B^+ \to \rho^+ \rho^0$. But in experimental side,
          the world average results of these decays
          have not  so much difference. Neither QCD factorization approach nor
          naive factorization  can explain this small difference.
          The similar situation  also appears in the decays $B \to
          \pi\pi$  \cite{hep-ph/0308039 , hep-ph/9411308}.  Many
          people have tried to explain this puzzle  \cite{hep-ph/0411373,hep-ph/0503077}.
          But for the $B\to \rho \rho$ decays, it is still early,
          since the very small branching ratio of $B^0 \to \rho^0\rho^0$ by
          experiments contradicts with isospin symmetry. We have
          to wait for the experiments.
   \item  From the Table~\ref{results}, we know that longitudinal polarization
          is dominant in decay $B^0 \to \rho^+ \rho^- $, $B^+ \to \rho^+
          \rho^0$ and $B^+ \to \rho^+\omega$, which occupies more than $90\%$
          contribution, and is  consistent with experimental data.  These
          results are also consistent with the prediction in naive
          factorization \cite{hep-ph/9804363}, because the transverse parts are $r_{\rho}^2$
          suppressed, where $r_{\rho}=m_{\rho}/m_B$.
          But for $B^0 \to \rho ^0 \rho ^0$ decay, the tree  emission diagrams
          are mostly cancelled in the Wilson coefficients. As we will see later in Table
          \ref{contribution}, the most important contributions for this decay are from
          the non-factorizable tree diagrams in Fig.\ref{fig1}(c) and (d) and also the
          penguin diagrams.
          With an additional gluon, the transverse polarization in
          the non-factorizable diagrams
          does not encounter helicity flip  suppression. The
          transverse polarization is at the same order as longitudinal polarization, which
          can also be seen in the column (c) and (d) of Table~\ref{results}.
          This scenario is new from the mechanism of the recently penguin dominant process
          $B\to \phi K^*$ \cite{hep-ph/0411305}, where the penguin annihilation guides the
          dominant transverse contribution. In fact, the $B^0 \to\omega
          \omega$ decay has a little larger longitudinal fraction
         is just due to the fact that there is no non-factorizable emission tree
         contribution for this decay in isospin symmetry.
\end{itemize}
\begin{table}[htb]
\caption{Contribution from different parts in $B^0 \to \rho^+
\rho^-$ and $B^0 \to \rho ^0 \rho ^0$: full contribution in  line
$ (1)$, ignore annihilation contribution in  line $ (2)$, without
 penguin operators in line $ (3)$, and without non-factorization
diagrams in  line $ (4)$.} \label{contribution}
\begin{center}
\begin{tabular}[t]{|r|c|c|c|c|c|c}
\hline $B^0 \to \rho^+
\rho^-$ & BR$ (10^{-6})$ & $f_L (\%)$ & $f_{\parallel }(\%)$ & $f_{\perp} (\%)$  \\
\hline
 (1)~~~~~~~&    35   & 94    & 3     & 3 \\
 (2)~~~~~~~&    35   & 94    & 3     & 3 \\
 (3)~~~~~~~&    32   & 94    & 3     & 3 \\
 (4)~~~~~~~&    38   & 96    & 2     & 2 \\
\hline\hline
$B^0 \to \rho ^0 \rho ^0$ & BR$(10^{-6})$ &$f_L (\%)$ & $f_{\parallel }(\%)$ & $f_{\perp} (\%)$  \\
\hline
(1)~~~~~~~&  0.94   &  60   & 22    & 18  \\
(2)~~~~~~~&  0.38   &  42   & 26    & 32  \\
(3)~~~~~~~&  0.25   &  18   & 41    & 41  \\
(4)~~~~~~~&  1.18   &  83  &  8.5    & 8.5  \\
\hline
\end{tabular}
\end{center}
\end{table}

Now we turn to discuss the contribution of different diagrams,
where  $B^0 \to \rho^+ \rho^- $ and $B^0 \to \rho ^0 \rho ^0$ are
taken as an example. In the Table~\ref{contribution}, we consider
full contribution in line $ (1)$, ignore annihilation contribution
in line $ (2)$, without all penguin operator in line $ (3)$, and
without non-factorization diagram in  line $ (4)$. From this
table, we can see that neither annihilation diagrams nor
non-factorizable diagrams can change the polarization fraction in
decay $B^0 \to \rho^+ \rho^- $. They only take about $4\%$
contribution in this decay mode just because the emission diagram
occupy very large part of the  contribution, which can also be
seen from Table~\ref{ratio}. However, the penguin operators
especially in annihilation diagrams play an important role in
decay $B^0 \to \rho ^0 \rho ^0$.

Of course, the final state interaction  is very important in
non-leptonic $B$ decays. They  can give $\mathcal{O} (10^{-6})$
corrections \cite{hep-ph/0409317}, but this can not change the
branching ratios much for  decay modes $B^0 \to \rho^+ \rho^-$ and
$B^+ \to \rho^+ \rho^0$ at order $10^{-5}$. Thus, in these two
decay modes, the final state interaction may not be important.
However, in decay $B^0 \to \rho ^0 \rho ^0$, the final state
interaction may afford larger contribution than our calculation
($10^{-7} -   10^{-6}$), that's to say, our perturbative part may
not be the dominant contribution. Although probably important, the
hadronic effects are not intensively discussed in this paper,
since they are beyond the topics of our PQCD approach. The
contributions of these two sides can be determined by experiments.

\section{CP Violation in $B^0 \to \rho^+ \rho^- $ and $B^+ \to
\rho^+ \rho^0(\omega)$}\label{sc:CP}

Studying CP violation is an important task in $B$ physics. In this
section, we discuss the CP violation in $B^0 \to \rho^+ \rho^- $
and $B^+ \to \rho^+ \rho^0(\omega)$ decays. The uncertainty in
$B^0 \to \rho^0 \rho^0 (\omega), \omega\omega $  for branching
ratios is so large that we will not discuss their CP violation
here, though it is also very important. In decay modes $B^0 \to
\rho^+ \rho^- $ and $B^+ \to \rho^+ \rho^0(\omega)$, longitudinal
part occupy nearly $95\%$ contribution. So we will neglect the
transverse parts in the following discussions.

\begin{figure}[htb]
\begin{center}
\includegraphics[scale=0.75]{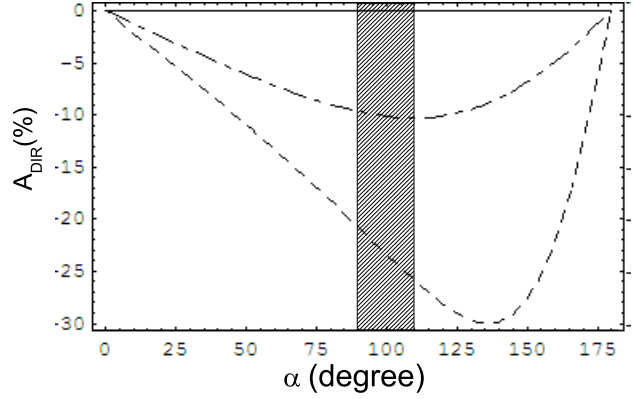}
\caption{Direct CP violation parameter $A^{dir}_{CP}$ as a
function of $\alpha$ with $\omega_B=0.4$. The solid line is for
$B^+ \to \rho^+ \rho^0$, dot-dashed for $B^0 \to \rho^+ \rho^- $
and dashed line for $B^+ \to \rho^+ \omega $. The shadow part is a
band with $90^\circ < \alpha <110^\circ$.} \label{Adirect}
\end{center}
\end{figure}

Using Eqs.~(\ref{M_H},\ref{M_H_2}), the direct CP violating
parameter is easily derived as a function of  CKM angle $\alpha$.
\begin{eqnarray}\label{adircp1}
A^{dir}_{CP}&=&\frac{{|M^+|}^2-{|M^-|}^2}{{|M^+|}^2+{|M^-|}^2}\nonumber
\\
&=&\frac{-2z\sin\alpha \sin\delta_L}{1+2z_L\cos\alpha
\cos\delta_L+z_L^2},
\end{eqnarray}
which is shown in Fig.\ref{Adirect}. The direct CP asymmetry is
about $(-10 \pm 4)\%$ in decay $B^0 \to \rho^+ \rho^- $. However,
the direct CP in decay $B^+ \to \rho^+ \rho^0$ is almost zero,
because there is no QCD penguin contribution while electroweak
penguin contribution is rather negligible. On the other hand,
because of large penguin contribution, the direct CP in $B^+ \to
\rho^+ \omega$ is about $(-23 \pm 7) \%$, which is even larger
than $B^0 \to \rho^+ \rho^- $. The uncertainty in above results
come from $90^\circ < \alpha <110^\circ$ and $0.3 < \omega_B <0.4$
in $B$ meson wave function.

\begin{figure}[htb]
\begin{center}
\includegraphics[scale=0.75]{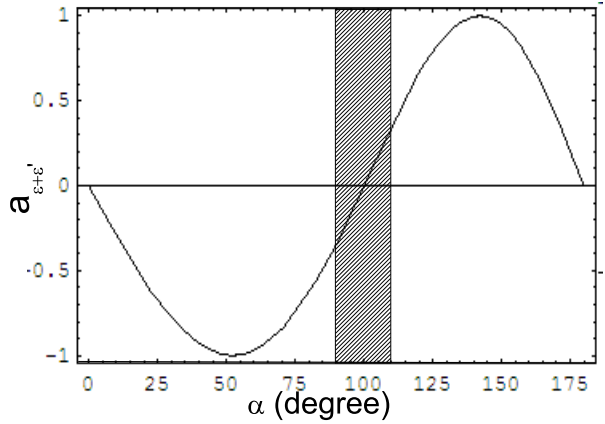}
\caption{Mixing induced CP violation parameters
$a_{\epsilon+\epsilon^{\prime}}$
   of $B^0 \to \rho^+ \rho^- $ as a function of CKM
angle $\alpha$ with $\omega=0.4$. The shadow part is a band with
$90^\circ < \alpha <110^\circ$.} \label{Aeta}
\end{center}
\end{figure}

For the neutral $B^0$ decays, there is more complication from the
$B^0-\bar B^0$ mixing. The time dependence of CP asymmetry is:
\begin{eqnarray}\label{adircp2}
A_{CP} &\simeq& A^{dir}_{CP} \cos (\Delta m t) +\sin (\Delta m t)
a_{\epsilon+\epsilon^{\prime}},
\end{eqnarray}
where $\Delta m$ is the mass difference between the two mass
eigenstates of neutral $B$ mesons. The  $A^{dir}_{CP}$ is already
defined in Eq.(\ref{adircp1}), while the mixing-related CP
violation parameter is defined as
\begin{eqnarray}
a_{\epsilon+\epsilon^{\prime}}=\frac{-2 \mathrm{Im}
(\lambda_{CP})}{1+|\lambda_{CP}|^2},
\end{eqnarray}
where
\begin{eqnarray}
\lambda_{CP}=\frac{V_{tb}^*V_{td}<f|H_{eff}|\bar
B|>}{V_{tb}V_{td}^*<f|H_{eff}|B>}.\label{lambdacp}
\end{eqnarray}
Using Eqs.~(\ref{M_H},\ref{M_H_2}), we derive as:
\begin{eqnarray}
\lambda_{CP}\simeq e^{2i\alpha}\frac{1+z_Le^{i
(\delta_L-\alpha)}}{1+z_Le^{i (\delta_L+\alpha)}}.
\end{eqnarray}
Thus, the parameter $a_{\epsilon+\epsilon^{\prime}}$ is a function
of $\alpha$, if the penguin pollution is very small,
$a_{\epsilon+\epsilon^{\prime}}$ is about $-\sin2\alpha$. From the
function relation of Fig.~\ref{Aeta}, we can see that
$a_{\epsilon+\epsilon^{\prime}}$ is not exactly equal to
$-\sin2\alpha$, because of the penguin pollution.

If we integrate the time variable t of Eq.(\ref{adircp2}), we will
get the total CP asymmetry as
\begin{eqnarray}
A_{CP}=\frac{1}{1+x^2}A_{CP}^{dir}+\frac{x}{1+x^2}a_{\epsilon+\epsilon^{\prime}}
\end{eqnarray}
with $x=\Delta m/\Gamma\simeq0.771$ for the $B^0-\bar B^0$ mixing
in SM  \cite{pdg}. Through calculating, we notice that the
$A_{CP}$ is $(-10\pm 4)\%$ with uncertainty from $90^\circ <
\alpha <110^\circ$ and $0.3 < \omega_B <0.4$.

\section{Summary}\label{sc:sum}
In this work, we calculate the branching ratios, polarizations and
CP asymmetry of $B\to\rho (\omega)\rho (\omega)$ decays in
 perturbative QCD approach based on $k_T$ factorization. After calculating all
diagrams, including non-factorizable diagrams and annihilation
diagrams, we found the branching ratios of $B^0 \to \rho^+ \rho^-
$ and $B^+ \to \rho^+ \rho^0$ are at order of $\mathcal{O}
(10^{-5})$, and the longitudinal contributions are more than
$95\%$. These results agree with the BaBar's data well. Moreover,
we also predict the direct CP violation in $B^0 \to \rho^+ \rho^-
$ and $B^+ \to \rho^+ \rho^0$, and mixing CP violation in $B^0 \to
\rho^+ \rho^- $, which may be important in extraction for the
angle $\alpha$. The longitudinal polarization for $B^0 \to
\rho^0\rho^0$, $\rho^0\omega$, $\omega\omega$ are suppressed to
$60\%-80\%$ due to the large non-factorizable tree contribution to
these decays. These results can be tested in $B$ factories in
future.

\section*{Acknowledgments}

This work is partly supported by the National Science Foundation
of China. We thank  C.-H. Chen for useful discussion and G.-L.
Song and D.-S. Du for reading the manuscript.


\end{document}